# Comparative Evaluation of 3D Reconstruction Methods for Immersive Visualization of Laboratory Objects

**Brian De La Cruz[1‡], Aaron Y. Zhao[2‡], Maitrey Gramopadhye[3], Sawyer J. Lazar[4], Xianming Tan[5], Daniel Szafir[6], and David S. Lawrence[7*]**

[1] Department of Computer Science, College of Arts and Sciences, The University of North Carolina, Chapel Hill, NC, USA, 27599; brian_dela_cruz@outlook.com
[2] Department of Computer Science, College of Arts and Sciences, The University of North Carolina, Chapel Hill, NC, USA, 27599; aaron.y.zhao1@gmail.com
[3] Department of Computer Science, College of Arts and Sciences, The University of North Carolina, Chapel Hill, NC, USA, 27599; maitrey@cs.unc.edu
[4] Department of Chemistry, College of Arts and Sciences, The University of North Carolina, Chapel Hill, NC, USA, 27599; slazar@unc.edu
[5] Department of Biostatistics, School of Public Health and the Lineberger Comprehensive Cancer Center, The University of North Carolina, Chapel Hill, NC, USA, 27599; xianming@email.unc.edu
[6] Department of Computer Science, College of Arts and Sciences, The University of North Carolina, Chapel Hill, NC, USA, 27599; dszafir@cs.unc.edu
[7] Division of Chemical Biology and Medicinal Chemistry, UNC Eshelman School of Pharmacy, Department of Chemistry, College of Arts and Sciences, Department of Pharmacology, UNC School of Medicine, The University of North Carolina, Chapel Hill, NC, USA, 27599; lawrencd@email.unc.edu
‡ These authors contributed equally to this study.
* Correspondence: lawrencd@email.unc.edu

**Abstract**

In this study, we examined whether current 3D reconstruction methods can support the creation of realistic holographic representations of laboratory objects for educational use. In this regard, we compared four approaches: photogrammetry, a neural radiance field (NeRF)–based method, Gaussian splatting, and LiDAR. These methods were used to generate holographic models of common laboratory items and their fidelity was evaluated by graduate students. Participants assessed the models for shape, color, texture, and visual defects using a repeated-measures design. Across objects, the NeRF-based method produced the most consistently high-fidelity representations, particularly for transparent, reflective, or low-texture items that were difficult to capture with other approaches. Shape and color were generally reproduced more successfully than texture, suggesting that some visual properties remain more challenging to represent accurately in educational holograms. Beyond identifying the strengths and limitations of each reconstruction method, the study demonstrates a practical workflow for creating immersive learning objects that may support pre-laboratory preparation, spatial reasoning, and student engagement in AR/MR-based educational environments. These findings offer design-relevant insights for educators and researchers developing immersive digital learning experiences.

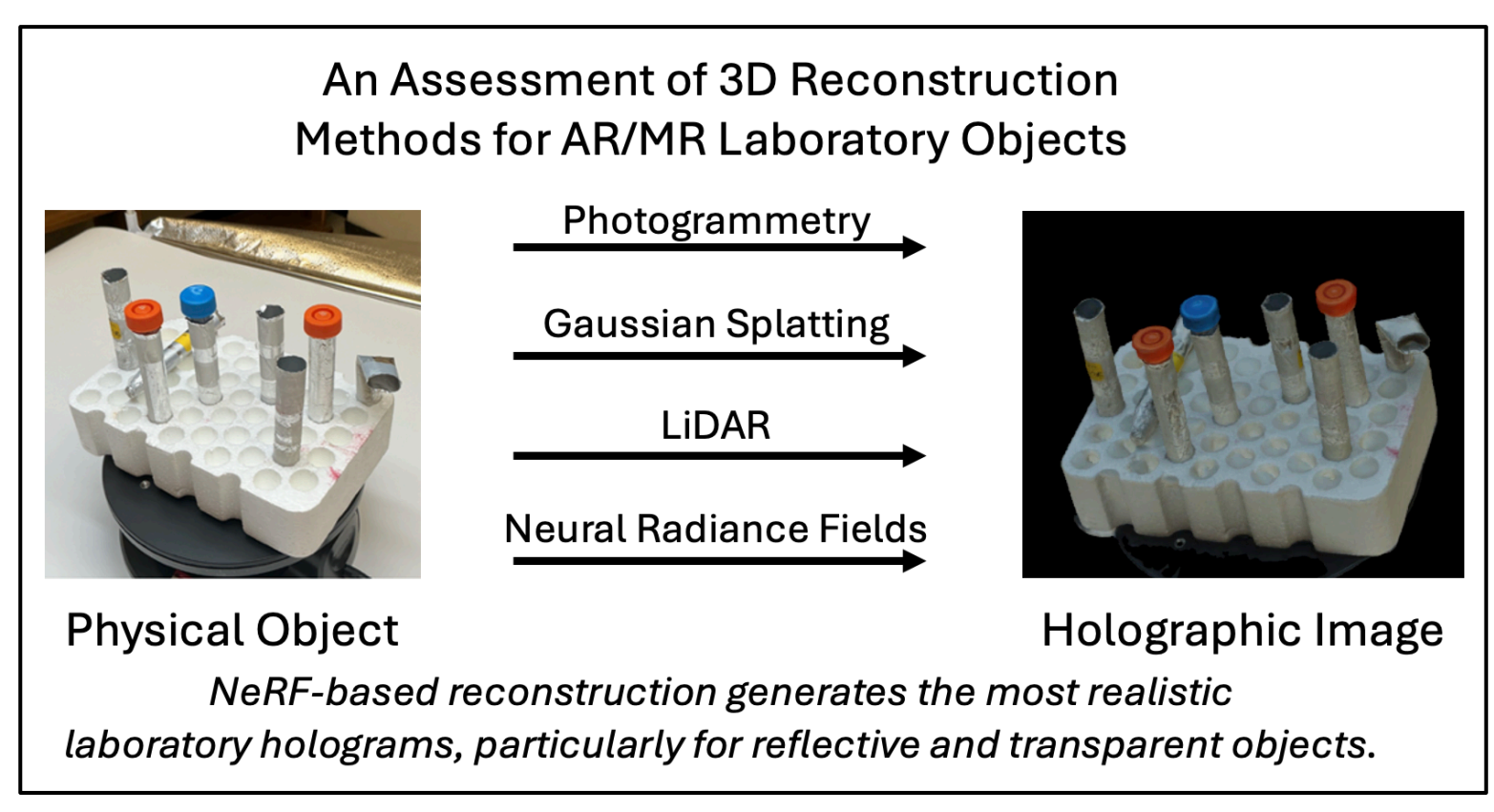

## 1. Introduction

Digital technologies are increasingly being used to expand what is possible in teaching and learning, particularly in contexts where learners benefit from seeing, manipulating, and reasoning about complex three-dimensional structures.[1] Among these technologies, augmented reality (AR) and mixed reality (MR) are especially promising because they place digital objects into the learner's physical environment while preserving spatial relationships, scale, and orientation.[2,3] This combination of realism and interactivity has made AR/MR attractive for educational applications ranging from science and engineering to health professions and technical training.[4-8] For learning tasks that depend on spatial understanding, the ability to view digital objects in context offer advantages over static images, videos, or conventional screen-based 3D models.[9]

A growing body of educational research suggests that immersive technologies can support engagement, motivation, and conceptual understanding when they are designed with clear pedagogical goals.[1] In particular, AR/MR environments may help learners develop spatial reasoning by allowing them to inspect objects from multiple viewpoints and connect abstract representations to physical experience. These attributes are relevant not only to molecular visualization or instrument training, but also to broader forms of laboratory and workplace learning in which learners must recognize equipment, understand layout, and anticipate how objects relate to one another in space. However, the educational value of AR/MR depends heavily on the quality of the digital objects being presented. If a virtual object is poorly rendered, lacks visual realism, or fails to preserve important features such as shape and texture, it may reduce learner trust, limit engagement, or weaken the instructional value of the experience.

Creating realistic digital objects for AR/MR remains a practical challenge. One route is to reconstruct physical objects using 3D scanning methods that convert images or sensor data into viewable models. Recent advances have made these methods more accessible, including smartphone-based tools that can generate 3D representations without specialized equipment.[10-12] Yet not all reconstruction methods perform equally well across object types. Some approaches are better suited to objects with strong geometric features, while others may better handle reflective, transparent, or low-texture surfaces. For educational applications, this distinction matters because learning environments often include a wide range of items with different visual and structural properties. A method that works well for one type of object may fail for another, which limits its usefulness for building reusable instructional content.

This technical variability has important educational implications, yet relatively little educational research has compared current methods in terms of their suitability for creating immersive learning materials. Most studies have focused on technical performance or on narrow use cases within a single discipline. Less attention has been given to how reconstruction quality affects the usability and educational potential of AR/MR learning objects across a broader set of materials. This gap is notable because educators need practical guidance about which methods are most appropriate for generating realistic digital resources that can support teaching, preparation, and exploration.

In this study, we examined four 3D reconstruction approaches: photogrammetry,[13] a neural radiance field (NeRF)–based method,[14] Gaussian splatting,[15] and LiDAR[16]. We've compared and contrasted how well these four methods support the creation of holographic objects. A range of common laboratory items were selected that vary in transparency, reflectivity, texture, and geometric complexity. Graduate student volunteers evaluated the resulting holograms for shape, color, texture, and defects. Our goal was not only to compare technical reconstruction quality, but also to identify which methods are most promising for producing realistic digital objects that could be used in immersive learning environments.

## 2. Materials and Methods

### 2.1. Imaging of Laboratory Equipment and Supplies.

12 common laboratory objects were selected that varied in visual and structural complexity. The set included items with different combinations of transparency, reflectivity, texture, color, and shape complexity, allowing us to examine how object characteristics influence the quality of the reconstructed models. These objects were chosen because they represent the range of objects learners may encounter in laboratory preparation and instruction. In addition, the provided a useful test of whether current reconstruction methods can support realistic educational representations.

The objects selected for scanning were an acetone safety polyethylene wash bottle (**1**), a glass 500 mL filter flask (**2**), a heating mantle (**3**), a 4 L amber acetonitrile bottle (**4**), Kimwipes™ (**5**), a laboratory jack (**6**), a cork ring (**7**), a 4 L plastic Nalgene™ beaker (**8**), a Bunsen burner (**9**), a StyrofoamTM test tube rack containing various objects (**10**), a mortar and pestle (**11**), and a mortar and pestle treated with a cyclododecane-based spray to provide a matte finish (**12**) (**Figure 1**). All objects were scanned on a Revopoint Dual Axis Turntable placed inside a Fasonic 32″ × 32″ studio light box. The light box was used to eliminate shadows and ensure consistent illumination.

Objects were scanned using Apple's LiDAR-enabled iPad Pro (3rd generation). A variety of smartphone-based 3D scanning apps are now commercially available.[17-20] We employed the KIRI Engine app given its ability employ four distinct technologies to capture, render, and model objects in 3D.

(1) Photogrammetry creates 3D structures from a series of approximately 200 two-dimensional photographs taken of an object from different angles and heights. Image output is constructed from a low-density, sparse 3D data set that emphasizes key object features such as vertices, edges, and faces. Although photogrammetry is known for its geometric accuracy, it has been reported to struggle with highly reflective or low-texture objects.[21]

(2) Gaussian splatting renders each of the millions of captured points in an object as a Gaussian distribution of characteristics, including position, color, size, and opacity, which are then projected onto a screen. Although the basic concept dates to the early 1990s, its first application to 3D object and scene rendering was reported in 2023.[15] Like photogrammetry, Gaussian splatting relies on 2D images acquired from multiple viewpoints. Unlike photogrammetry, however, it blends individual points so that missing information can be interpolated, which may introduce geometric inaccuracies.[22]

(3) LiDAR emits laser pulses toward an object and uses the return time to calculate distances. The resulting point cloud is a direct measure of object geometry. Accordingly, LiDAR is well suited for reproducing structure, but it is less effective for capturing color and fine texture.[23]

(4) Neural radiance fields were first described in 2021 and, like the other methods used here, require scene capture from multiple positions.[14] A radiance field maps the interaction of light, color, and density with individual elements of an object as a function of position and viewing angle, and the object representation is then constructed through neural network training. NeRFs are often described as especially effective for highly reflective, low-texture objects. A current limitation of NeRFs is their high computational demand, which becomes especially relevant when rendering large scenes.[24] We employed a NeRF variant known as Featureless Object Scan developed by KIRI Engine.[25]

Although each of these methods has reported strengths and weaknesses, an important question is how well those strengths overcome inherent limitations when applied to laboratory objects.

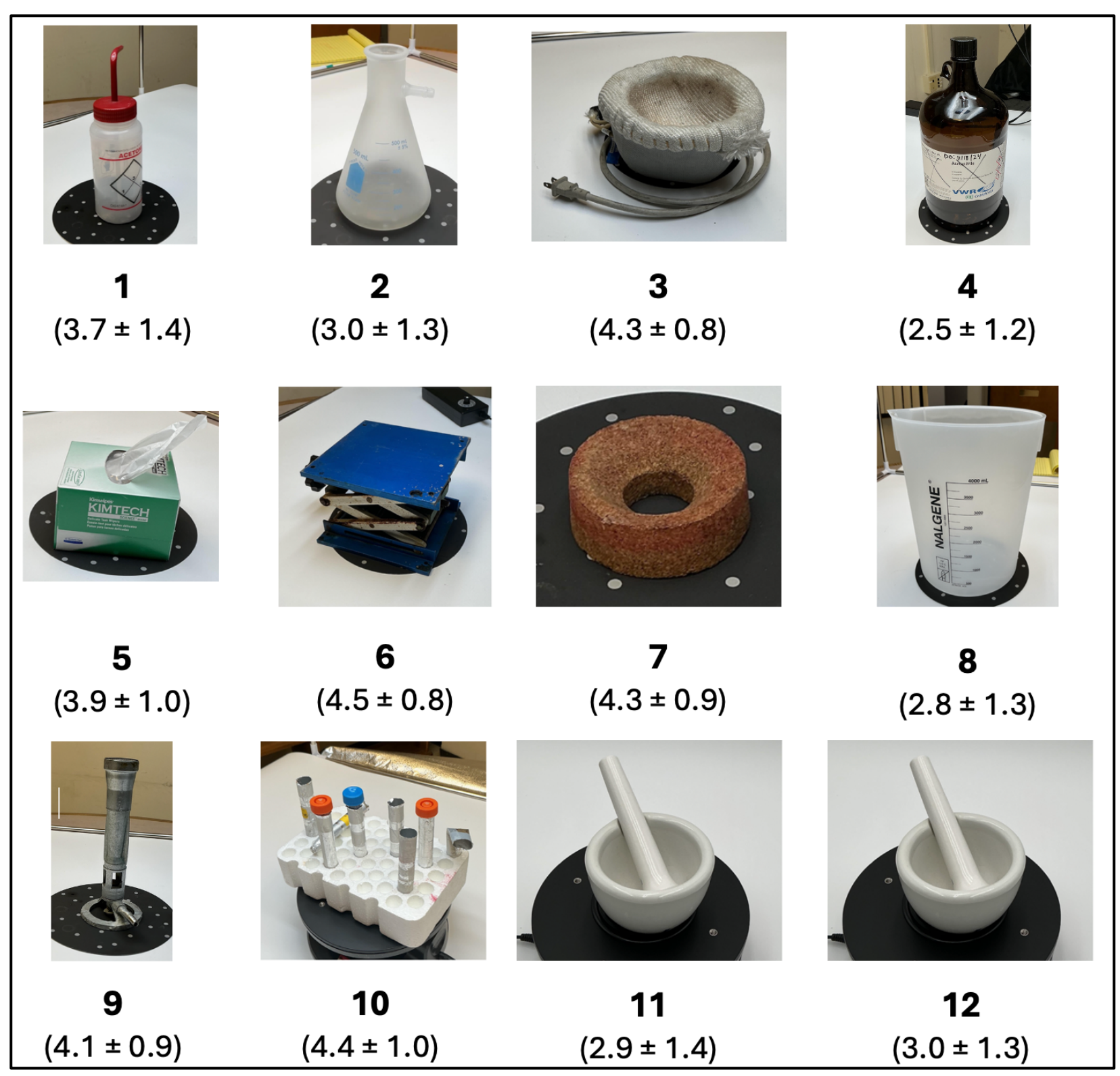

**Figure 1**. The objects chosen for scanning include an acetone polyethylene wash bottle (**1**), a glass 500 mL filter flask pretreated with a cyclododecane-based spray (**2**), a heating mantle (**3**), a 4 L amber acetonitrile bottle (**4**), Kimwipes™ (**5**), a laboratory jack (**6**), a cork ring (**7**), a 4 L plastic Nalgene™ beaker (**8**), a Bunsen burner (**9**), a Styrofoam™ test tube rack containing various objects (**10**), a mortar and pestle (**11**), and a mortar and pestle treated with a cyclododecane-based spray (**12**). Values in parentheses are a summary of averaged scores for all the imaging methods (photogrammetry, Gaussian splatting, LiDAR, and NeRF) and all the physical traits (shape, texture, color, and defects) for each object, where 1 = worst and 5 = best. n = 276 data points (17x4x4) generated from 17 student assessors.

**2.2. Acquisition and Transformation of 3D Models into a HoloLens Viewing Experience.**

The Unity Game Engine was employed to edit and format the 48 (12 objects x 4 scanning methods) digital models for viewing in an augmented reality headset (HoloLens). The exported models from KIRI Engine were positioned around the world origin, and each model was rotated to align with the world axis, providing a fixed 3D coordinate system and ensuring consistent orientation across models. This step ensured that all models were displayed in a comparable format for participant evaluation. The goal was to create a realistic and stable viewing experience that would allow users to inspect each object from multiple perspectives in an AR/MR environment. Given the stream heavy nature of the Unity application, a PC-driven HoloLens format was employed so that the PC's central and graphics processing units were used instead of the less robust HoloLens' mobile chips.

**2.3. Assessment of the 3D Holographic Models.**

Graduate student volunteers from the Department of Chemistry and the School of Pharmacy participated as expert raters. These participants were selected because they had familiarity with laboratory objects and could provide informed judgments about the realism and educational usefulness of the holographic models. All participants evaluated every model, enabling repeated-measures comparison across methods.

Participants received a brief introduction to the AR/MR viewing system and were shown how to inspect the holographic objects (Supporting Information, **Figure S-1**). They then independently reviewed the holographic objects through the HoloLens head-mounted display. Each of the 12 objects were presented as a group of 4, representing the 4 distinct scanning methods. All 48 objects (12 objects x 4 scanning methods) were simultaneously displayed (**Figure 2**). Each of the 48 holographic images were assessed in terms of shape, color, texture, and absence of defects.

Shape: how accurately the overall form of the object is reproduced.

Color: how well the object's color is represented.

Texture: how effectively surface detail and material appearance are captured.

Defects: the extent to which the model is free of visible artifacts or distortions.

These attributes were chosen because they reflect key qualities that influence the educational realism of AR/MR learning objects.[26]

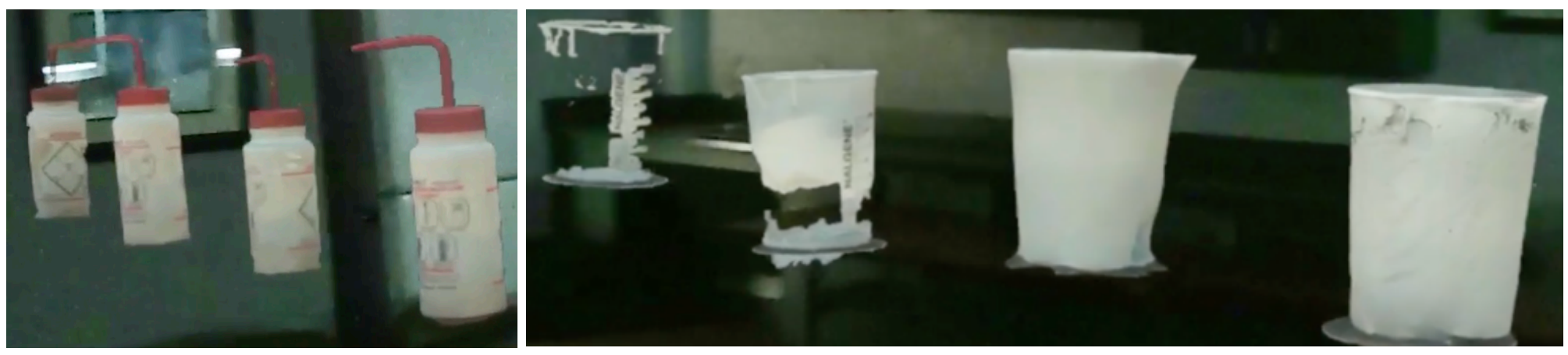

**Figure 2.** Image capture, through the HoloLens, of the acetone polyethylene wash bottle **Object 1** (left) and the Nalgene™ plastic beaker **Object 8** (right). The holographic images were acquired (from left to right) via photogrammetry, Gaussian splatting, LiDAR, and NeRF, respectively. The poor quality of the beaker model (especially those generated via photogrammetry and Gaussian splatting) is evident. Note: images and videos captured off the HoloLens head-mounted display, are of poorer quality relative to the user's visual experience. This is due to hardware constraints of the HoloLens photo/video camera, the lower frame rate, loss of resolution due to resource sharing, and positioning of the camera lens above the user's eyes (which results in warping of the hologram).[27] The visual experience of the user is best exemplified by the corresponding images directly captured off of the PC (**Figures 3 - 5**).

Ratings for shape, color, texture, and absence of defects were recorded, via an iPAD (Qualtrics survey) using a 1-to-5 scale, where higher scores indicated more realistic and instructionally useful representations (Table S-13). The evaluation was designed to capture both visual fidelity and the aspects of realism most likely to matter in educational settings.

**2.4. Data analysis.**

Each participant rated all four reconstruction methods. Consequently, we analyzed the data using within-subject statistical methods. For each object and visual attribute, we compared the four methods using nonparametric tests appropriate for repeated measures. When overall differences were detected, we conducted pairwise comparisons with correction for multiple testing. We also summarized performance across objects by averaging ratings within each method and across visual attributes. This approach allowed us to identify both method-level trends and object-specific patterns relevant to educational use.

For each object and visual attribute, differences among the four scanning methods were assessed using the Friedman test,[28] which is appropriate for repeated-measures data and does not assume normally distributed responses. When the global Friedman test was significant, *post hoc* paired comparisons between methods were conducted using Wilcoxon signed-rank tests with correction for multiple comparisons. Descriptive summaries are reported as mean, median, and standard deviation for each method-object-attribute combination. We also calculated the average of the four attribute scores for each method to compare overall method performance across objects. All tests were two-sided, and multiplicity was controlled using the Benjamini–Hochberg false discovery rate procedure[29] for Friedman tests across objects and visual attributes and Holm's method[30] for *post hoc* paired comparisons. Adjusted p-values < 0.05 were considered statistically significant. All analyses were performed in R 4.4.0.21.[31]

## 3. Results

### 3.1. Object 1. Acetone Polyethylene Wash Bottle.

The acetone polyethylene wash bottle combines a variety of features that can potentially draw on both the strengths and weaknesses of each of the four imaging method. The bottle is characterized by well-defined writing and color, and good shape definition, but with a slight opacity and modest reflective properties (**Figure 3a**).

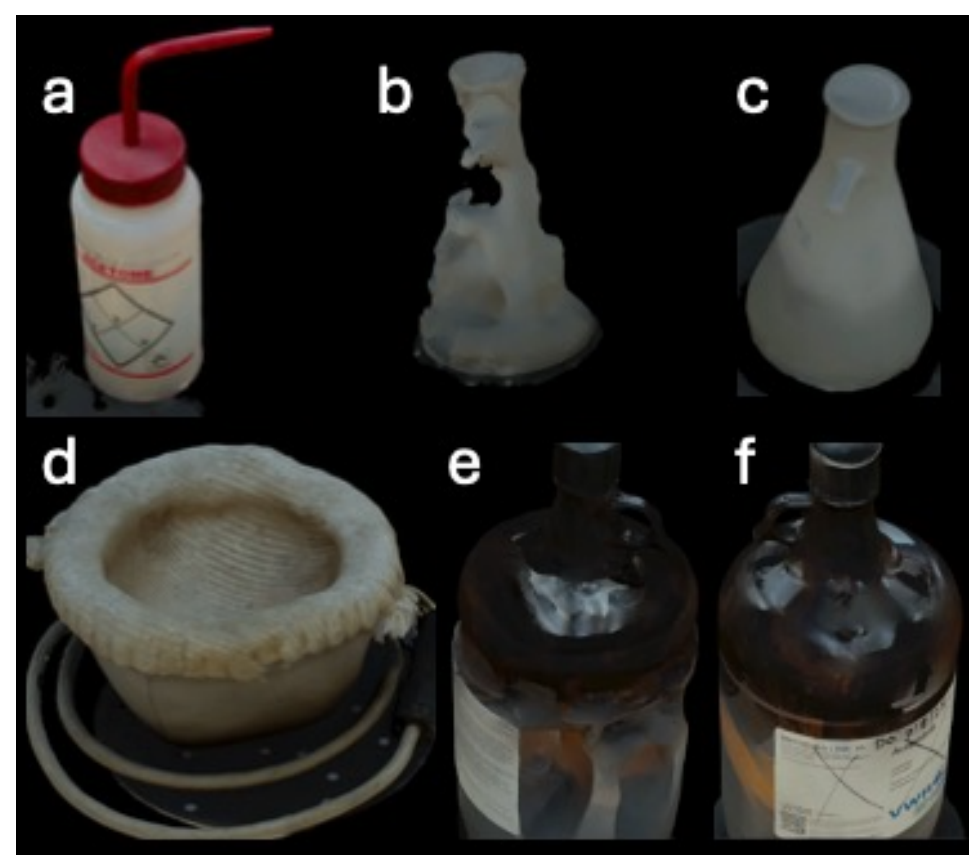

**Figure 3.** Holograms **1 – 4.** (**a)** The NeRF-generated model of **Object 1** (NeRF) displays excellent shape, color, texture, and is mostly free of defects. (**b** - **c**) Models generated for both the unsprayed (not shown) and sprayed glass filter flask (**Object 2**) were generally unacceptable, as exemplified by the photogrammetry-derived hologram (**b**). However, the shape of the NeRF-generated model (**c**) is excellent as is the absence of defects. (**d**) All methods furnished excellent models for the **Object 3** heating mantle (NeRF model shown in **d**). (**e** – **f**). In general, 3D imaging of the amber glass bottle (**Object 4**) resulted in numerous defects [see the lower half of the Gaussian splatting model in (**e**)]. However, NeRF furnished a model free of defects and excellent shape replication. Note: images were lightened by ~20% using Apple's Preview app for clarity purposes but are otherwise unaltered. Images were directly captured off of the PC, rather than using the on-board HoloLens camera.

All four methods furnished excellent color reproduction (Table 1 and Table S-1, Supplementary Materials). By contrast, both LiDAR and photogrammetry struggled with replicating the shape of the spray bottle, which is surprising given the supposed strengths of these methods in terms of furnishing geometric accuracy. We suspect that the reflective nature of the object is responsible for the inability to precisely capture shape. In addition, photogrammetry performed poorly with respect to texture and the overall model has a number of defects. Statistical analysis confirmed these qualitative statements. Specifically, method differences are significant for shape, texture, and defects (Friedman tests, BH-adjusted-p $< 0.05$) but not for color (BH-adjusted-p $= 0.06$) (Supporting Information, Table S-13). *Post hoc* paired comparisons (with Holm's adjustment, Supporting Information, Table S-14) between methods support the qualitative assessment that photogrammetry and LiDAR underperform whereas Gaussian splatting and NeRF produced higher-fidelity models (in terms of fewer defects). A brief qualitative summary for all the objects, using the analysis discussed for **Object 1**, is provided in **3.2 – 3.11** (see data outlined in Tables S-13 and S-14, Supporting Information).

### 3.2. Object 2. Glass 500 mL Filter Flask.

All four methods performed so poorly on the transparent, highly reflective object that we decided not to share these models with graduate student volunteers. Instead, the filter flask was treated with a cyclododecane-based spray, which eliminates transparency and dramatically reduces reflectivity. In spite of this pre-treatment, both photogrammetry (**Figure 3b**) and LiDAR were unable to generate effective 3D models (Table 1 and Table S-2, Supplementary Materials). NeRF furnished the only structurally well-defined model (**Figure 3c**). All four methods supplied unexceptional reproductions of the blue color present on the flask, which is presumably due to interference by the spray.

**Table 1.** Assessed fidelity of the virtual reproduction of each object's combined traits (shape, texture, color, defects) by the four different scanning methods. n = 68 data points (17 x 4) generated from 17 student assessors. The best and most reliable (SE) value for each object is highlighted in boldface. Values comparable to the best/reliable scores are italicized.

| Object | Photogrammetry | Gaussian | LiDAR | NeRF |
|---|---|---|---|---|
| **1** | 2.6 ± 1.6 | 4.2 ± 0.8 | 3.4 ± 1.1 | **4.6 ± 0.6** |
| **2*** | 1.9 ± 1.2 | 3.0 ± 0.9 | 2.8 ± 1.0 | **4.3 ± 0.9** |
| **3** | *4.3 ± 0.8* | *4.1 ± 0.9* | *4.4 ± 0.8* | **4.4 ± 0.7** |
| **4** | 2.2 ± 1.0 | 1.8 ± 0.9 | 2.4 ± 1.0 | **3.7 ± 1.1** |
| **5** | 3.8 ± 0.9 | 3.8 ± 1.1 | 3.6 ± 1.1 | **4.5 ± 0.7** |
| **6** | *4.5 ± 0.9* | *4.4 ± 0.8* | **4.5 ± 0.7** | **4.5 ± 0.7** |
| **7** | **4.4 ± 0.8** | 4.0 ± 1.0 | *4.4 ± 0.9* | *4.3 ± 0.9* |
| **8** | 1.9 ± 1.3 | 2.7 ± 1.2 | **3.3 ± 1.0** | 3.3 ± 1.2 |
| **9** | **4.3 ± 0.9** | 2.9 ± 1.0 | **4.2 ± 0.8** | 4.0 ± 1.0 |
| **10** | **4.6 ± 0.9** | *4.3 ± 1.0* | *4.4 ± 1.0* | *4.3 ± 1.0* |
| **11** | 2.8 ± 1.3 | 2.4 ± 1.4 | **3.6 ± 1.1** | 2.8 ± 1.4 |
| **12*** | 3.3 ± 1.1 | 2.1 ± 1.2 | **3.8 ± 0.9** | 3.1 ± 1.0 |

*The filter flask (**Object 2** ) and motor and pestle (**Object 12**) were sprayed with cyclododecane to create a temporary, sublimatable, matte finish to reduce the object's reflectivity.

### 3.3. Object 3. Heating Mantle.

In addition to low reflectivity, the chosen object is well-worn containing color variations and a tear in the fabric. All four methods generated excellent holographic replicas (**Figure 3d**) (Table 1 and Table S-3, Supplementary Materials).

### 3.4. Object 4. 4 L Amber Acetonitrile Bottle.

Like the filter flask, the glass acetonitrile bottle is highly reflective, but unlike the flask, it is not transparent. Photogrammetry, Gaussian splatting (**Figure 3e**), and LiDAR were unable to generate reasonable defect-free models (Table 1 and Table S-4, Supplementary Materials). By contrast, NeRF furnished a good fidelity model, most notably excelling in capturing the overall shape of the bottle with few defects (**Figure 3f**).

### 3.5. Object 5. Kimwipes™.

The green colored box contains a modest hint of reflectivity along with a single, geometrically complex, Kimwipe. All four methods performed well with respect to capturing the color of the object and, to a modestly lesser extent, the shape and texture (Table 1 and Table S-5, Supplementary Materials). However, NeRF once again excelled in terms of shape and producing a defect-free model (note the Kimwipe defects in **Figure 4a** versus that of **Figure 4b**).

### 3.6. Object 6. Laboratory Jack.

The laboratory jack contains a smooth medium blue plate, slightly rusted metal supports, with overall low reflectivity. All four scanning methods produced excellent 3D models (**Figure 4c**) (Table 1 and Table S-6, Supplementary Materials).

### 3.7. Object 7. Cork Ring.

This highly textured, speckled-colored object was well reproduced by all methods, with little or no defects. There was, however, a slight fall off, relative to some of the best models, in capturing the overall texture and color of the cork ring, suggesting that high complexity is difficult to precisely capture with current methods (**Figure 4d**) (Table 1 and Table S-7, Supplementary Materials).

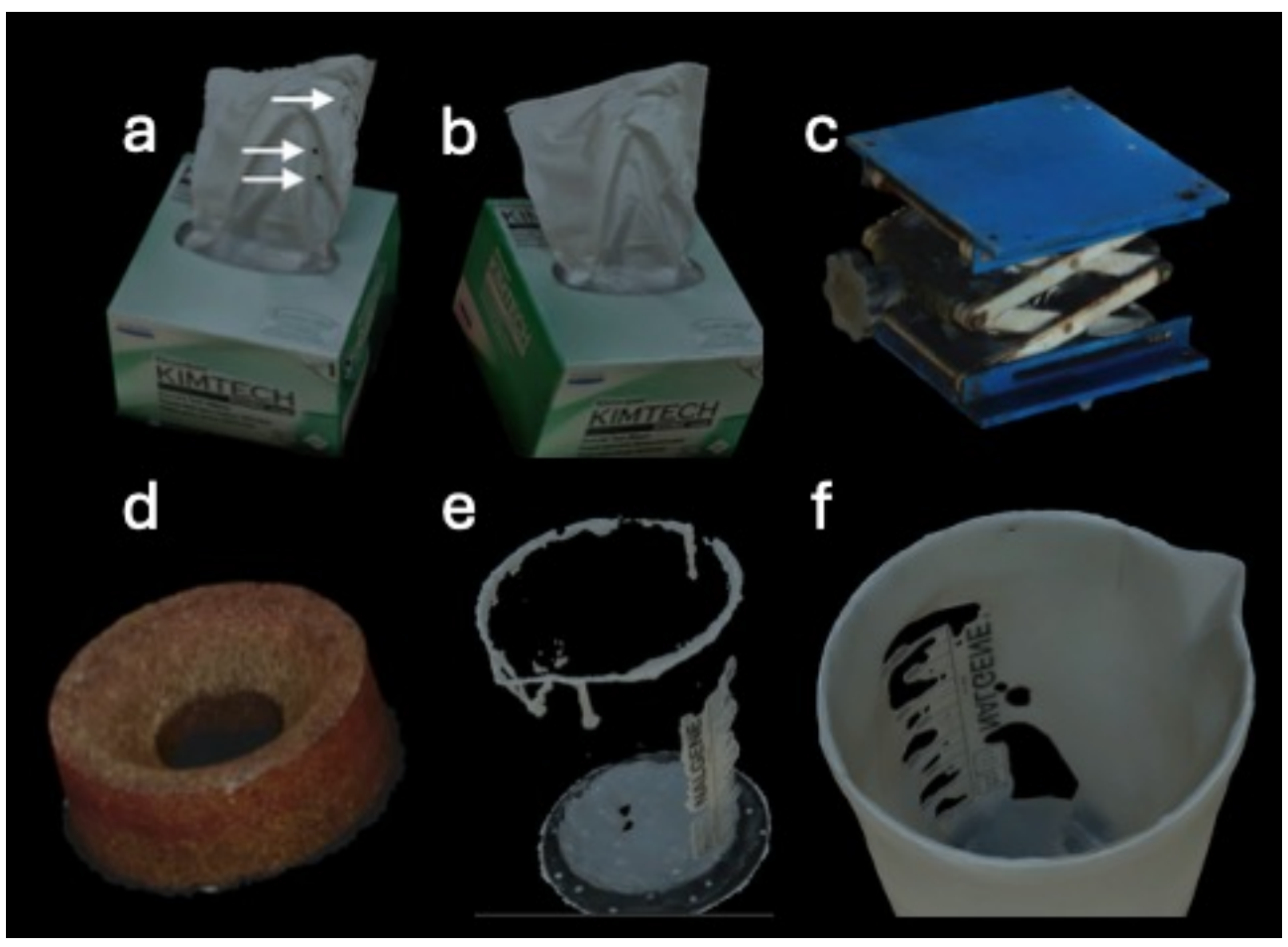

**Figure 4.** Holograms **5 – 8.** (**a**). The photogrammetry-generated model of **Object 5** displays excellent shape, color, and texture, but a few minor defects in the Kimwipe™ (arrows) are present. (**b**) The corresponding NeRF model is free of defects. (**c**) All 3D scanning methods furnished excellent models of **Object 6**, the blue colored laboratory jack (LiDAR model is shown). (**d**). All 3D scanning methods generated excellent models of **Object 7**, the highly textured cork ring (photogrammetry model is shown). (**e**) All scanning methods experienced significant challenges in rendering the plastic Nalgene™ beaker, as exemplified by the photogrammetry model. (**f**) The LiDAR model of **Object 7** proved to be the best, but even in this case significant defects are evident. Note: images were lightened by ~20% using Apple's Preview app for clarity purposes but are otherwise unaltered. Images were directly captured off of the PC, rather than using the on-board HoloLens camera.

**3.8. Object 8. 4 L Plastic Nalgene™ Beaker**.

This glassware variant is semi-transparent and moderately reflective. None of the scanning methods were able to produce high-fidelity models (**Figure 4e**), with LiDAR (**Figure 4f**) and NeRF displaying the best (albeit modest) results (Table 1 and Table S-8, Supplementary Materials).

**3.9. Object 9. Bunsen Burner**.

This metal object is non-transparent with a dull, non-homogenous finish. Good models, with relatively few defects, were generated by all methods (**Figure 5a**) (Table 1 and Table S-9, Supplementary Materials). Interestingly, the quality of texture capture is somewhat diminished relative to the best results obtained for other objects, which may be consistent with challenges associated with reproducing complex surfaces.

**3.10. Object 10. Styrofoam™ Test Tube Rack Containing Various Objects**.

Given the apparent difficulties associated with reproducing features that are complex (high textured, speckled-colored) we decided to examine an object with structural and color complexity at a more macro-level. This object contains, in addition to Styrofoam™, aluminum foil, orange and blue caps, and different oriented geometries. All of the 3D scanning methods provided excellent holographic replicas (**Figure 5b**) (Table 1 and Table S-10, Supplementary Materials).

**3.11. Objects 11 and 12. Unsprayed (11) and Cyclododecane-Sprayed (12) Mortar and Pestel.**

This object is smooth with few features and highly reflective. Curiously, the only method which performed adequately was LiDAR (**Figure 5c**) (Table 1 and Tables S-11 and S-12, Supplementary Materials). By contrast, Gaussian splatting was by far the worst performer (**Figure 5d**). Furthermore, there are only subtle differences in the quality of the models created from the unsprayed (**Figures 5c** and **5d**) and sprayed objects (**Figures 5e** and **5f**).

**Table 1.** Assessed fidelity of the virtual reproduction of the four individual traits of each object for the combined scanning methods (photogrammetry, Gaussian splatting, LiDAR, and NeRF). n = 68 data points (17 x 4) generated from 17 student assessors. The best and most reliable (SE) value for each object is highlighted in boldface.

| Object | Photogrammetry | Gaussian | LiDAR | NeRF |
|---|---|---|---|---|
| **1** | 3.6 ± 1.3 | 3.6 ± 1.3 | **4.5 ± 0.6** | 3.1 ± 1.6 |
| **2*** | 3.1 ± 1.4 | 2.7 ± 1.1 | **3.4 ± 1.2** | 2.8 ± 1.5 |
| **3** | *4.4 ± 0.8* | 4.2 ± 0.8 | 4.2 ± 0.9 | **4.5 ± 0.7** |
| **4** | **2.9 ± 1.5** | *2.5 ± 1.0* | **2.5 ± 0.9** | 2.1 ± 1.3 |
| **5** | 4.2 ± 1.0 | 3.7 ± 1.1 | **4.3 ± 0.7** | 3.5 ± 1.1 |
| **6** | **4.6 ± 0.7** | *4.4 ± 0.8* | **4.5 ± 0.6** | *4.4 ± 0.9* |
| **7** | *4.5 ± 0.8* | 4.0 ± 1.0 | 3.9 ± 1.0 | **4.6 ± 0.8** |
| **8** | 2.9 ± 1.4 | 2.6 ± 1.3 | **3.4 ± 1.2** | 2.2 ± 1.0 |
| **9** | **4.4 ± 0.8** | 3.9 ± 1.1 | 4.0 ± 0.9 | *4.2 ± 0.9* |
| **10** | **4.7 ± 0.7** | 4.0 ± 1.1 | *4.4 ± 0.8* | *4.4 ± 1.0* |
| **11** | 3.2 ± 1.4 | 2.5 ± 1.3 | **3.6 ± 1.2** | 2.4 ± 1.2 |
| **12*** | 3.2 ± 1.4 | 2.7 ± 1.1 | **3.6 ± 1.0** | 2.8 ± 1.2 |

*The filter flask (**Object 2** ) and motor and pestle (**Object 12**) were sprayed with cyclododecane to create a temporary, sublimatable, matte finish to reduce the object's reflectivity.

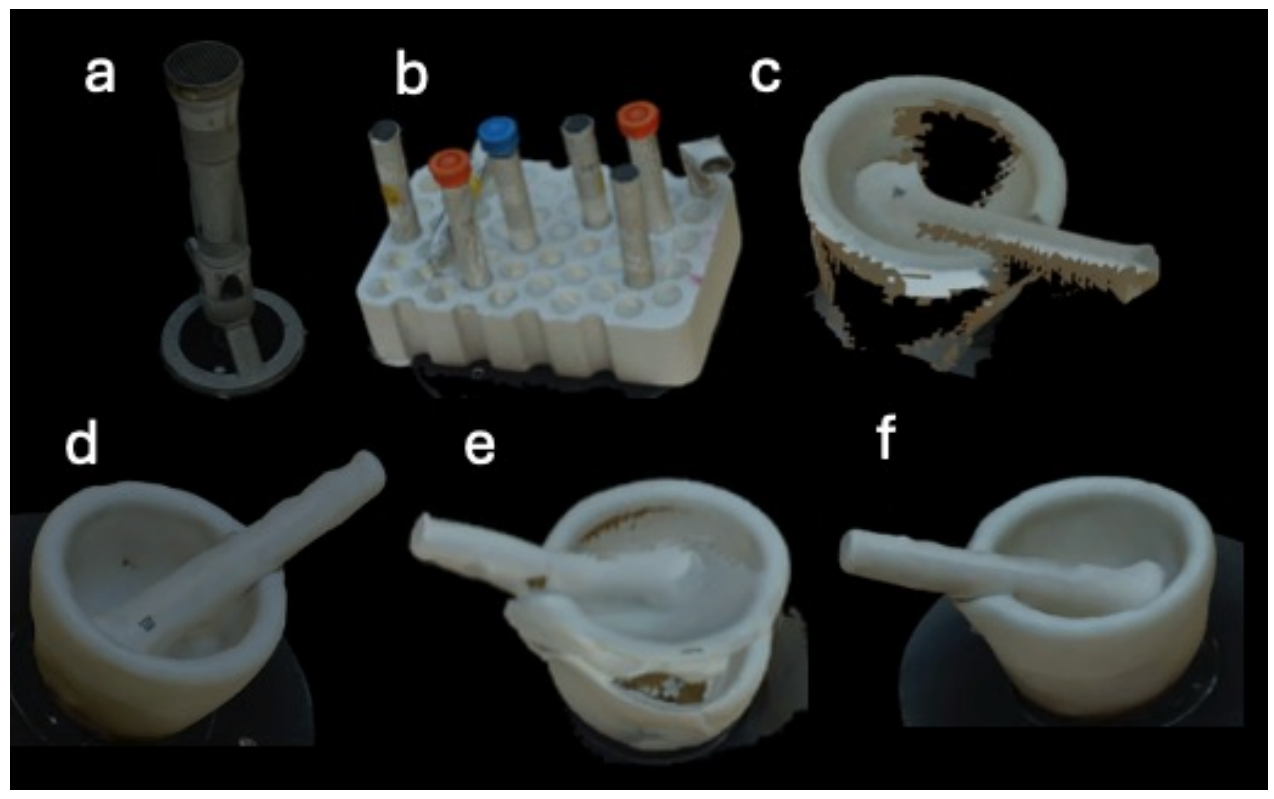

**Figure 5.** Holograms **9 – 12.** (**a** - **b**). The photogrammetry-generated models of **Objects 9** and **10** display excellent shape, color, and texture, with minimal defects. (**c** - **f**) The mortar and pestle models are of generally poor quality. The most deficient of these, (**c**, unsprayed) and (**e**, sprayed) were generated by Gaussian splatting. However, even the "best" models (**d**, unsprayed; **f**, sprayed), which were generated by LiDAR, display deformities, especially within the pestle. Note: images were lightened by ~20% using Apple's Preview app for clarity purposes but are otherwise unaltered. Images were directly captured off of the PC, rather than using the on-board HoloLens camera.

### 3.12. Method Performance.

Across the 12 objects, the NeRF-based method produced the most consistently high-fidelity holographic models. It generated the largest number of models with average scores above the high-fidelity threshold (arbitrarily set ≥ 4) and the fewest models in the fair-to-poor range (arbitrarily set ≤ 2.5). LiDAR also performed well for selected objects, whereas photogrammetry and Gaussian splatting more often produced lower-rated models.

### 3.13. Object-dependent Differences.

Method performance varied substantially by object type. The reconstruction methods performed best for objects with clear structural features and low visual complexity, such as the heating mantle (**Object 3**), laboratory jack (**Object 6**), Bunsen burner (**Object 9**), and test tube rack (**Object 10**). In contrast, transparent, reflective, or smooth featureless objects were more difficult to accurately reproduce. These included the glass filter flask (**Object 2**), amber bottle (**Object 4**), plastic beaker (**Object 8**), and mortar and pestle [without (**Object 11**) and with (**Object 12**) cyclododecane spray].

**3.14. Visual Attributes.**

Shape and color were generally reproduced more successfully than texture (**Table 2**). Texture was the most difficult attribute to capture consistently across methods and objects. Defects were also more common in models of transparent or reflective objects, particularly when using photogrammetry and Gaussian splatting.

**3.15. Statistical Comparisons.**

Repeated-measures analyses confirmed significant differences among reconstruction methods for several object-attribute combinations, especially for shape, texture, and defects. *Post hoc* comparisons revealed that NeRF and, in some cases, LiDAR outperformed photogrammetry and Gaussian splatting for challenging objects. For several simpler objects, however, all four methods produced comparably strong results.

**3.16. Educational Relevance.**

Participants reported that the holographic models were easy to view and assess, suggesting that the workflow is usable for immersive educational applications. The results indicate that current reconstruction methods can support realistic AR/MR representations of laboratory objects, but that fidelity depends strongly on object properties. This is especially relevant for pre-laboratory instruction, where students must learn to recognize equipment, anticipate spatial relationships, and prepare for safe laboratory work. In this respect, our findings align with prior work showing that virtual and immersive tools can help bridge the safety gap between lecture-based instruction and hands-on research laboratory experiences, particularly by allowing learners to encounter laboratory environments and procedures in a low-risk setting before entering the lab.[32] The present workflow extends that idea by demonstrating that realistic 3D digital surrogates of actual laboratory objects can be created and inspected from multiple viewpoints, potentially strengthening both safety preparation and spatial understanding in AR/MR-based learning environments.

## 4. Discussion

This study examined whether current 3D reconstruction methods can support the creation of realistic AR/MR learning objects. The results show that immersive representations of laboratory objects are feasible with existing tools, but that reconstruction quality depends strongly on the visual and structural properties of the object being scanned. In particular, the NeRF-based method produces the most consistent results across object types, suggesting that it may be especially useful for educational materials that include reflective, transparent, or low-texture items.

A key finding is that shape and color are reproduced more reliably than texture. This matters for education because learners often rely on shape[33] and color[34] to identify objects and understand their function, while texture contributes more subtly to realism and material perception[35]. The relative difficulty of capturing texture also suggests that current AR/MR learning objects may be most effective when instructional goals emphasize recognition, spatial orientation, and object identification rather than fine-grained surface detail.

This study also highlights an important design issue for educational technology researchers: the suitability of a reconstruction method depends on the intended learning context. Objects with simple geometry and low reflectivity are generally reproduced well by all four methods, whereas transparent or highly reflective objects can be challenging. This implies that educators and designers should match the reconstruction method to the instructional purpose and object type rather than assuming that one approach will work equally well in all cases. For example, methods that perform well with complex or reflective objects may be more appropriate for creating immersive pre-laboratory materials, while simpler methods may suffice for less demanding instructional tasks.

The workflow used in this study demonstrates a practical route for developing AR/MR learning objects that can be inspected from multiple viewpoints in a standardized format. These holographic images may find utility in pre-class preparation, guided exploration, and active learning activities that benefit from visual realism and spatial context. In addition, the positive response from the graduate student assessors suggests that the technology is sufficiently intuitive to be incorporated into educational settings without substantial training overhead.

Several features emerged that are directly relevant to teaching effectiveness. First, the technology proved to be highly intuitive; students required no more than a brief (≤5 min) introduction before independently engaging with the system. This low cognitive and logistical barrier implies that minimal class time is required, enabling rapid integration into instructional settings. Second, students efficiently evaluated all four characteristics across 48 objects in approximately 20 min without instructor intervention, indicating that the platform supports student-centered, self-directed exploration; an important component of effective active-learning environments. Third, informal verbal feedback was overwhelmingly positive, with students frequently expressing surprise and enthusiasm upon first interacting with the holographic images. This initial engagement was reinforced by consistently high agreement regarding the realism of the representations, suggesting that the platform can provide an accurate and immersive approximation of laboratory equipment. Such realism is critical for supporting observational learning, spatial reasoning, and conceptual transfer to physical laboratory settings. Finally, while current limitations related to highly transparent or reflective objects represent a design threshold for some applications, the technology is already well suited for simulating many objects in a laboratory environment, where visual familiarity and procedural context are central to student learning. Collectively, these findings indicate that the platform is intuitive, efficient, and engaging, with strong potential to enhance teaching effectiveness by improving accessibility to laboratory experiences, increasing student engagement, and supporting meaningful preparation for hands-on laboratory work.[36-39]

A potential application of this workflow is pre-laboratory AR/MR safety experiences for trainees entering a new research laboratory environment. Prior work using virtual reality has already shown that immersive simulation can strengthen laboratory safety education,[32] and AR/MR may extend this benefit by adding spatially grounded, real-world object placement and orientation. An AR/MR experience could help trainees learn to recognize both potential safety hazards and clear safety violations before they enter the laboratory, thereby addressing a critical gap in current safety training. Additional applications include procedural spatial reasoning, conceptual understanding of abstract processes, practice prior to hands-on work, and reducing dependence on expensive or limited physical resources. The present study serves as a foundation for these and other potential applications by demonstrating that realistic holographic laboratory objects can be created and inspected from multiple viewpoints in an intuitive format. Future work will build on this foundation by evaluating how these models support comprehension, retention, transfer, and training across different learner populations and instructional settings.

## 5. Conclusions

We've found that current 3D reconstruction methods can be used to create realistic AR/MR learning objects, but that their effectiveness depends on the visual properties of the object being represented. Among the methods tested, the NeRF-based technology produces the most consistent high-quality results, particularly for objects that are reflective, transparent, or difficult to capture. More broadly, the findings indicate that shape and color are more readily reproduced than texture, and that some object types remain challenging for all current reconstruction approaches.

These results suggest that immersive holographic objects may be especially useful for supporting spatial understanding, object recognition, and pre-laboratory preparation.[36-39] The workflow demonstrated here offers a practical way to generate digital learning materials that can be viewed in AR/MR environments and explored from multiple perspectives. Although the present study did not measure learning outcomes directly, it provides design-relevant evidence that can inform future development of immersive educational resources.

Future research should examine how these holographic objects affect student learning, engagement, and cognitive load in authentic instructional settings. It will also be important to determine which reconstruction methods are most appropriate for different educational goals and object types. Taken together, our results provide a foundation for using 3D reconstruction and AR/MR technologies to support more interactive, spatially rich forms of learning.

**Supplementary Materials:** The following supporting information can be downloaded at https://www.mdpi.com/article/doi/s1, Figure S-1, Qualtrics survey to assess quality of the 3D models (pdf); Tables S1 – S14, Shapes, Texture, Color, and Defects Associated with 3D Scanning Methods Photogrammetry; Gaussian Splatting; LiDAR; Neural Radiance Field (NeRF) Featureless Object Scan for Twelve Common Laboratory Objects; Table S-13. Friedman analysis of statistically meaningful differences for each object (1 - 12) and visual attribute (shape, texture, color, defects) amongst the four different methods (photogrammetry, Gaussian Splatting, LiDAR, NERF); Table S-14. Post hoc paired comparisons between methods were performed on those objects and attributes that displayed statistically significant differences using Friedman's test (Table S-13).

**Author Contributions:** All authors contributed to the study conception and design. Material preparation and data collection were performed by Brian De La Cruz, Aaron Y. Zhao, Maitrey Gramopadhye, and Sawyer J. Lazar. Statistical analysis was performed by Xianming Tan. The first draft of the manuscript was written by David S. Lawrence. All authors read and approved the final manuscript.

**Funding:** We gratefully acknowledge funding from the UNC Eshelman School of Pharmacy. This research was supported by the NSF under Award #2222953.

**Institutional Review Board Statement:** The experimental protocol was submitted to UNC's Institutional Review Board entitled "Augmented Reality: What are the best methods for imaging and virtually reproducing laboratory equipment?" (IRBIS 25-1830). A status of "no conflict" was determined.

**Informed Consent Statement:** "Informed consent was obtained from all subjects involved in the study. The student assessors furnished anonymous assessments of the holographic images.

**Data Availability Statement:** All supporting data for this study is either furnished in the manuscript or the supporting information, which is available online at https://www.mdpi.com/article/doi/s1.

**Acknowledgments:** We are pleased to acknowledge the contributions of the graduate student volunteers from the Department of Chemistry (College of Arts and Sciences) and the Division of Chemical Biology and Medicinal Chemistry (UNC School of Pharmacy) who scored the 3D models. We thank the Department of Chemistry for the use of an office to image the laboratory equipment and supplies and of a conference room to display the corresponding 3D virtual models.

**Conflicts of Interest:** The authors declare no conflicts of interest. The funders had no role in the design of the study; in the collection, analyses, or interpretation of data; in the writing of the manuscript; or in the decision to publish the results.

## Supplementary Material

# Comparative Evaluation of 3D Reconstruction Methods for Immersive Visualization of Laboratory Objects

**Table of Contents**

**Supporting Tables S1 - S12**. Shapes, Texture, Color, and Defects Associated with 3D Scanning Methods Photogrammetry; Gaussian Splatting; LiDAR; Neural Radiance Field (NeRF) Featureless Object Scan for Twelve Common Laboratory Objects.

*Note: green (≥4.0) and red (≤2.5); n = 17 for each method, n = 68 datapoints for All Methods and All Traits*

**Table S-1**. **Object 1**. Squirt Bottle

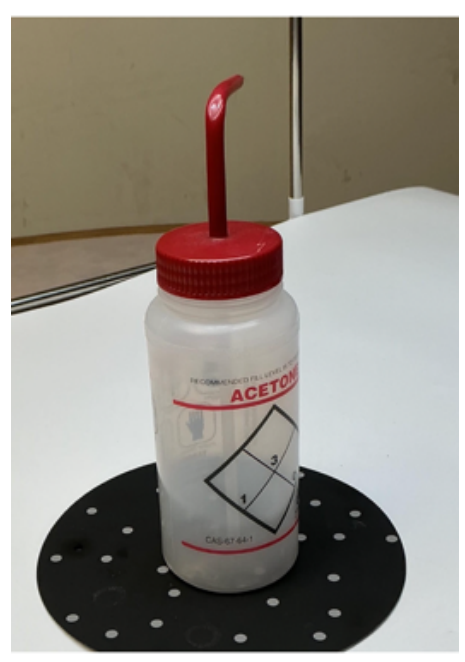

| Method | Shape | Texture | Color | Defects | All Traits |
|---|---|---|---|---|---|
| Photogram | 2.5 ± 1.3 | 2.4 ± 1.5 | 4.3 ± 0.8 | 1.0 ± 0.0 | 2.6 ± 1.6 |
| Gaussian | 4.2 ± 0.8 | 3.9 ± 1.0 | 4.5 ± 0.5 | 4.1 ± 1.0 | 4.2 ± 0.8 |
| LiDAR | 2.8 ± 0.9 | 3.7 ± 1.2 | 4.4 ± 0.7 | 2.7 ± 0.8 | 3.4 ± 1.1 |
| NeRF | 4.8 ± 0.6 | 4.3 ± 0.8 | 4.4 ± 0.7 | 4.7 ± 0.5 | 4.6 ± 0.6 |
| All Methods | 3.6 ± 1.3 | 3.6 ± 1.3 | 4.5 ± 0.6 | 3.1 ± 1.6 | |

**Table S-2**. **Object 2**. Sprayed Filter Flask

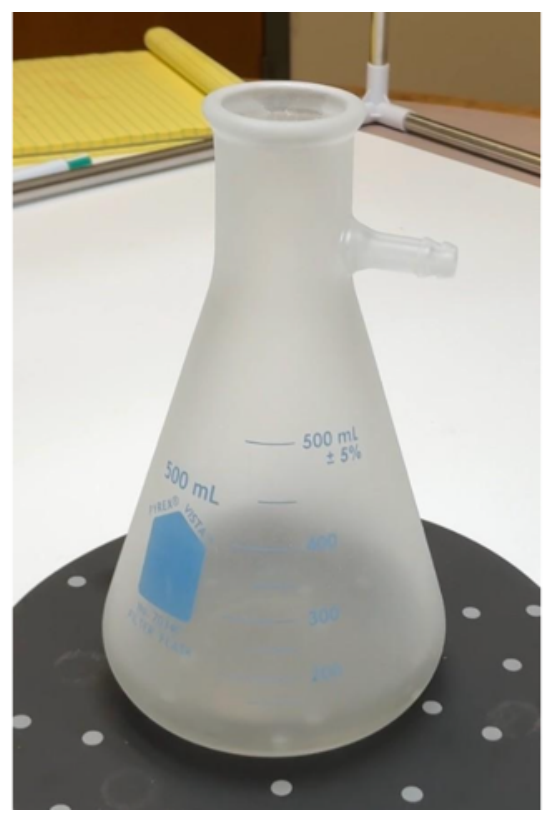

| Method | Shape | Texture | Color | Defects | All Traits |
|---|---|---|---|---|---|
| Photogram | 1.9 ± 1.3 | 1.6 ± 0.9 | 3.0 ± 1.3 | 1.0 ± 0.0 | 1.9 ± 1.2 |
| Gaussian | 3.2 ± 0.6 | 2.6 ± 0.8 | 3.5 ± 1.1 | 3.0 ± 0.7 | 3.0 ± 0.9 |
| LiDAR | 2.5 ± 0.9 | 2.6 ± 0.7 | 3.4 ± 1.2 | 2.6 ± 0.8 | 2.8 ± 1.0 |
| NeRF | 4.8 ± 0.4 | 3.8 ± 1.0 | 3.7 ± 1.1 | 4.8 ± 0.4 | 4.3 ± 0.9 |
| All Methods | 3.1 ± 1.4 | 2.7 ± 1.1 | 3.4 ± 1.2 | 2.8 ± 1.5 | |

**Table S-3**. **Object 3**. Heating Mantle

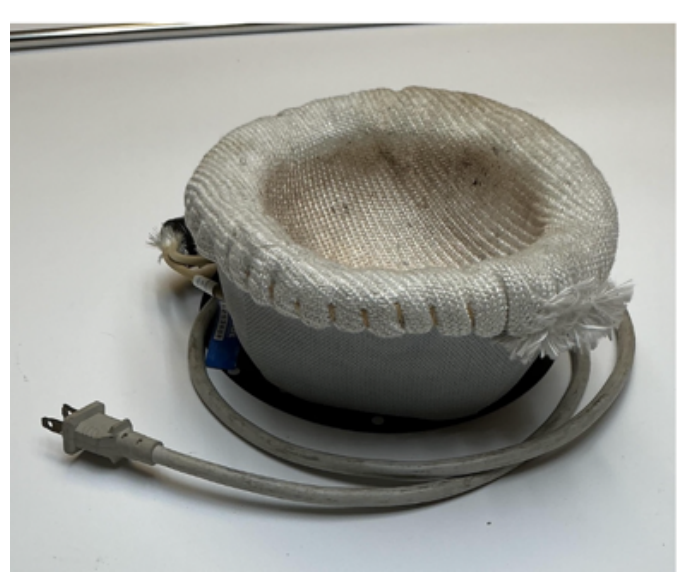

| Method | Shape | Texture | Color | Defects | All Traits |
|---|---|---|---|---|---|
| Photogram | 4.4 ± 0.9 | 4.3 ± 0.8 | 4.1 ± 0.9 | 4.5 ± 0.7 | 4.3 ± 0.8 |
| Gaussian | 4.2 ± 1.0 | 3.8 ± 1.0 | 4.2 ± 1.0 | 4.3 ± 0.7 | 4.1 ± 0.9 |
| LiDAR | 4.4 ± 0.7 | 4.5 ± 0.7 | 4.2 ± 1.0 | 4.6 ± 0.6 | 4.4 ± 0.8 |
| NeRF | 4.5 ± 0.6 | 4.2 ± 0.7 | 4.3 ± 1.0 | 4.6 ± 0.6 | 4.4 ± 0.7 |
| All Methods | 4.4 ± 0.8 | 4.2 ± 0.8 | 4.2 ± 0.9 | 4.5 ± 0.7 | |

**Table S-4**. **Object 4**. Amber Solvent Bottle

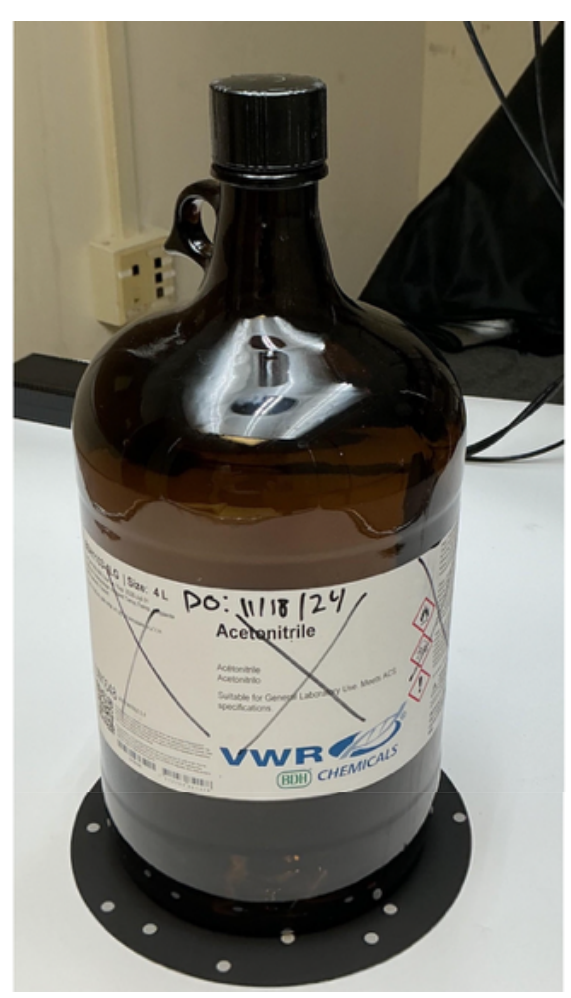

| Method | Shape | Texture | Color | Defects | All Traits |
|---|---|---|---|---|---|
| Photogram | 2.4 ± 1.3 | 2.3 ± 0.9 | 2.5 ± 0.9 | 1.4 ± 0.5 | 2.2 ± 1.0 |
| Gaussian | 1.8 ± 1.0 | 2.1 ± 0.9 | 2.2 ± 0.8 | 1.1 ± 0.2 | 1.8 ± 0.9 |
| LiDAR | 2.7 ± 1.1 | 2.5 ± 0.9 | 2.4 ± 0.9 | 2.1 ± 0.8 | 2.4 ± 1.0 |
| NeRF | 4.6 ± 0.6 | 3.2 ± 1.1 | 2.8 ± 0.8 | 4.1 ± 0.9 | 3.7 ± 1.1 |
| All Methods | 2.9 ± 1.5 | 2.5 ± 1.0 | 2.5 ± 0.9 | 2.1 ± 1.3 | |

**Table S-5**. **Object 5**. Kimwipes™ Box

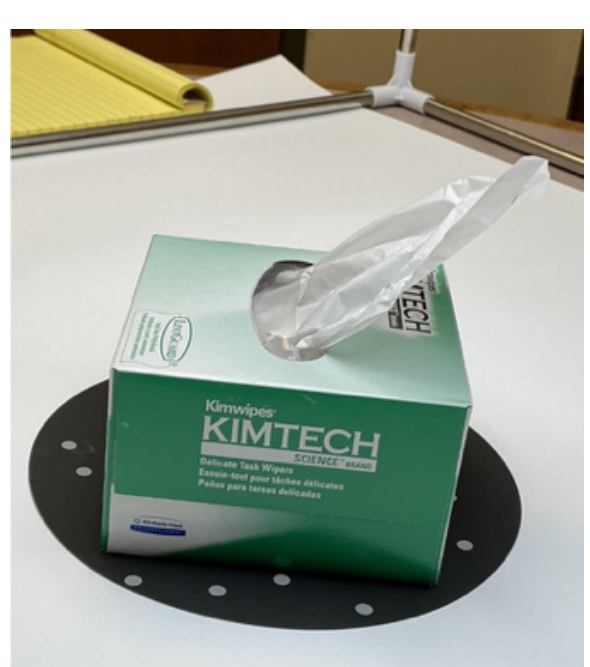

| Method | Shape | Texture | Color | Defects | All Traits |
|---|---|---|---|---|---|
| Photogram | 4.1 ± 1.1 | 3.8 ± 0.8 | 4.2 ± 0.6 | 3.3 ± 0.9 | 3.8 ± 0.9 |
| Gaussian | 4.1 ± 1.2 | 3.6 ± 1.1 | 4.2 ± 0.7 | 3.1 ± 1.0 | 3.8 ± 1.1 |
| LiDAR | 3.8 ± 0.9 | 3.2 ± 1.3 | 4.2 ± 0.7 | 3.2 ± 1.1 | 3.6 ± 1.1 |
| NeRF | 4.7 ± 0.5 | 4.3 ± 0.8 | 4.5 ± 0.7 | 4.6 ± 0.8 | 4.5 ± 0.7 |
| All Methods | 4.2 ± 1.0 | 3.7 ± 1.1 | 4.3 ± 0.7 | 3.5 ± 1.1 | |

**Table S-6**. **Object 6**. Laboratory Jack

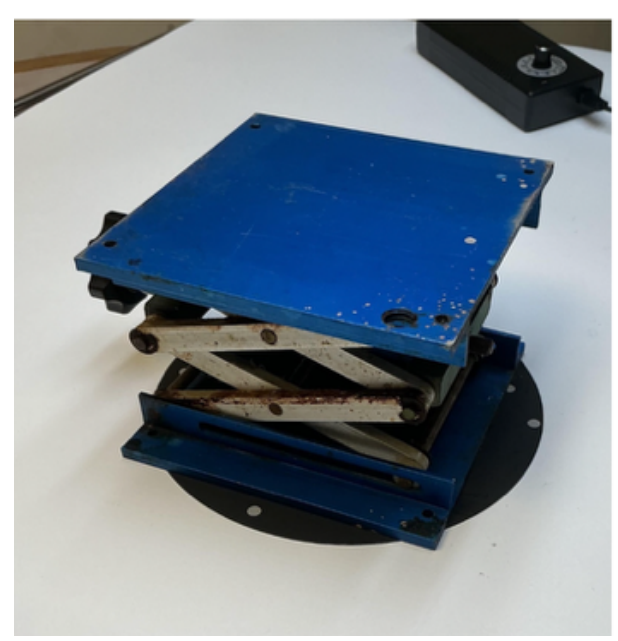

| Method | Shape | Texture | Color | Defects | All Traits |
|---|---|---|---|---|---|
| Photogram | 4.6 ± 0.9 | 4.5 ± 1.0 | 4.5 ± 0.7 | 4.3 ± 1.0 | 4.5 ± 0.9 |
| Gaussian | 4.7 ± 0.6 | 4.2 ± 1.0 | 4.4 ± 0.7 | 4.5 ± 0.8 | 4.4 ± 0.8 |
| LiDAR | 4.6 ± 0.8 | 4.5 ± 0.6 | 4.7 ± 0.5 | 4.5 ± 1.0 | 4.5 ± 0.7 |
| NeRF | 4.6 ± 0.6 | 4.4 ± 0.8 | 4.5 ± 0.6 | 4.5 ± 0.6 | 4.5 ± 0.7 |
| All Methods | 4.6 ± 0.7 | 4.4 ± 0.8 | 4.5 ± 0.6 | 4.4 ± 0.9 | |

**Table S-7**. **Object 7**. Cork Ring

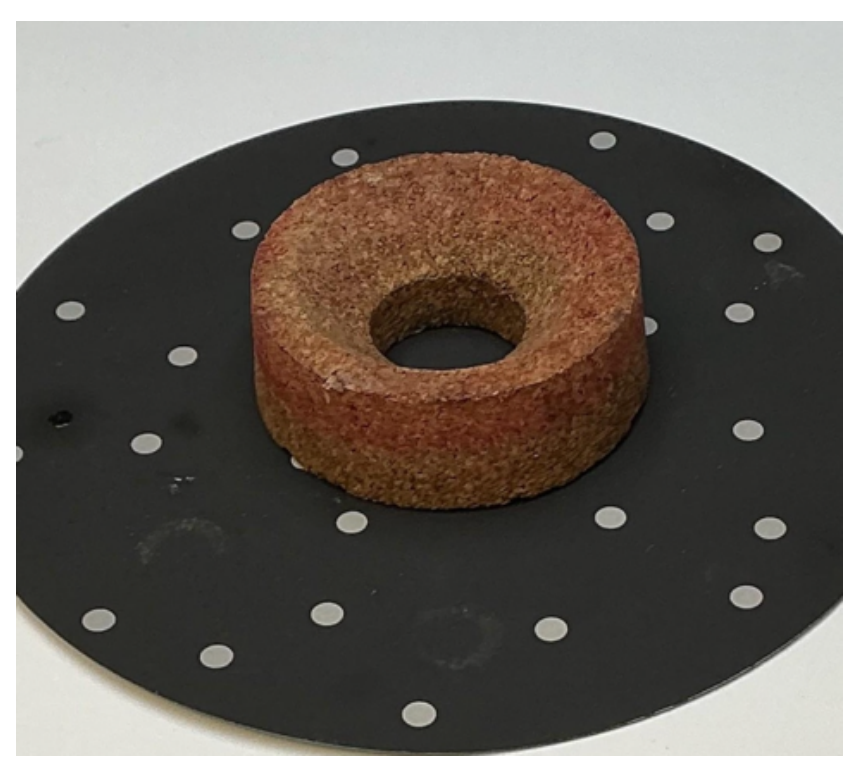

| Method | Shape | Texture | Color | Defects | All Traits |
|---|---|---|---|---|---|
| Photogram | 4.6 ± 0.6 | 4.4 ± 0.7 | 4.1 ± 1.0 | 4.6 ± 0.9 | 4.4 ± 0.8 |
| Gaussian | 4.3 ± 1.0 | 3.3 ± 1.0 | 3.7 ± 1.0 | 4.5 ± 0.8 | 4.0 ± 1.0 |
| LiDAR | 4.7 ± 0.8 | 4.3 ± 1.0 | 4.0 ± 1.0 | 4.7 ± 0.5 | 4.4 ± 0.9 |
| NeRF | 4.7 ± 0.8 | 4.0 ± 1.1 | 4.0 ± 1.0 | 4.5 ± 0.9 | 4.3 ± 0.9 |
| All Methods | 4.5 ± 0.8 | 4.0 ± 1.0 | 3.9 ± 1.0 | 4.6 ± 0.8 | |

**Table S-8. Object 8**. Plastic Nalgene™ Beaker

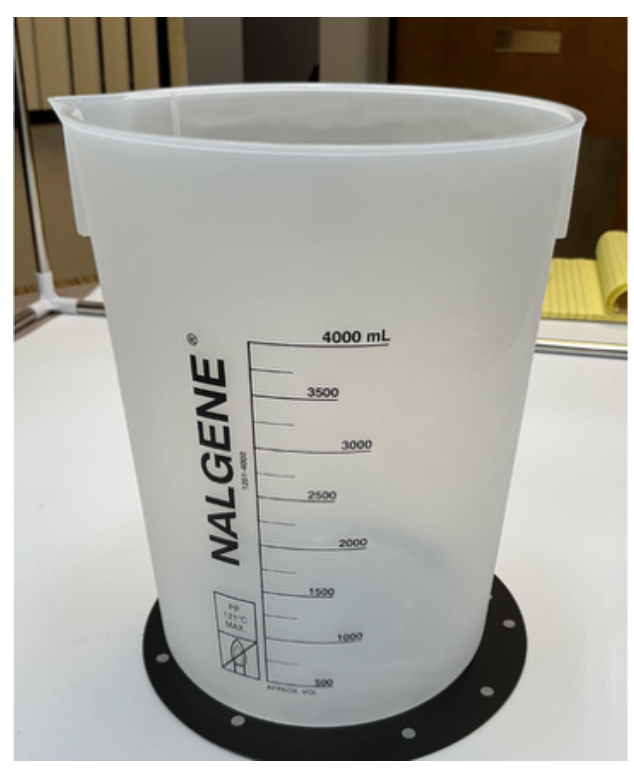

| Method | Shape | Texture | Color | Defects | All Traits |
|---|---|---|---|---|---|
| Photogram | 2.0 ± 1.4 | 1.6 ± 1.0 | 2.9 ± 1.3 | 1.1 ± 0.5 | 1.9 ± 1.3 |
| Gaussian | 2.8 ± 1.3 | 2.7 ± 1.2 | 3.5 ± 1.1 | 1.8 ± 0.4 | 2.7 ± 1.2 |
| LiDAR | 3.2 ± 1.0 | 3.4 ± 1.2 | 3.7 ± 1.1 | 2.8 ± 0.7 | 3.3 ± 1.0 |
| NeRF | 3.7 ± 1.4 | 2.8 ± 1.1 | 3.4 ± 1.1 | 3.1 ± 1.0 | 3.3 ± 1.2 |
| All Methods | 2.9 ± 1.4 | 2.6 ± 1.3 | 3.4 ± 1.2 | 2.2 ± 1.0 | |

**Table S-9. Object 9**. Bunsen Burner

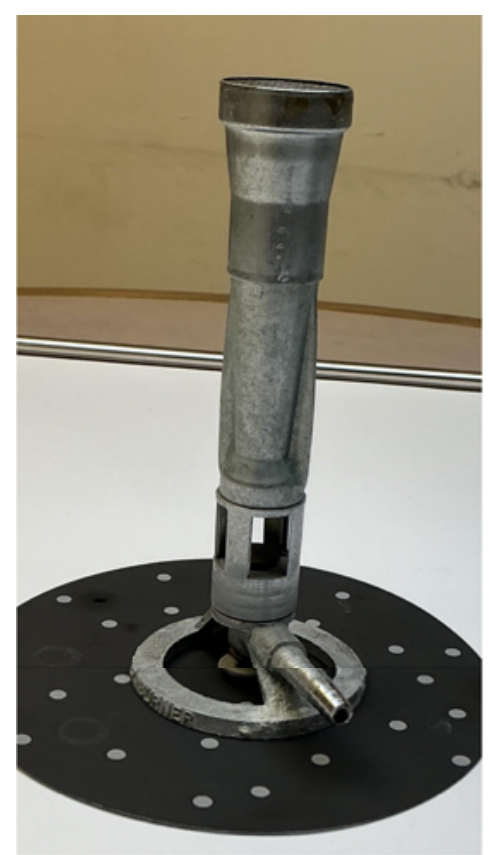

| Method | Shape | Texture | Color | Defects | All Traits |
|---|---|---|---|---|---|
| Photogram | 4.6 ± 0.7 | 4.2 ± 1.1 | 4.1 ± 1.0 | 4.4 ± 0.9 | 4.3 ± 0.9 |
| Gaussian | 4.2 ± 0.9 | 3.6 ± 1.1 | 3.9 ± 0.9 | 4.0 ± 1.0 | 3.9 ± 1.0 |
| LiDAR | 4.5 ± 0.8 | 4.0 ± 0.9 | 4.1 ± 0.8 | 4.2 ± 0.9 | 4.2 ± 0.8 |
| NeRF | 4.4 ± 0.9 | 3.8 ± 1.2 | 4.0 ± 0.9 | 4.4 ± 0.8 | 4.0 ± 1.0 |
| All Methods | 4.4 ± 0.8 | 3.9 ± 1.1 | 4.0 ± 0.9 | 4.2 ± 0.9 | |

**Table S-10**. **Object 10**. Styrofoam™ Test Tube Rack

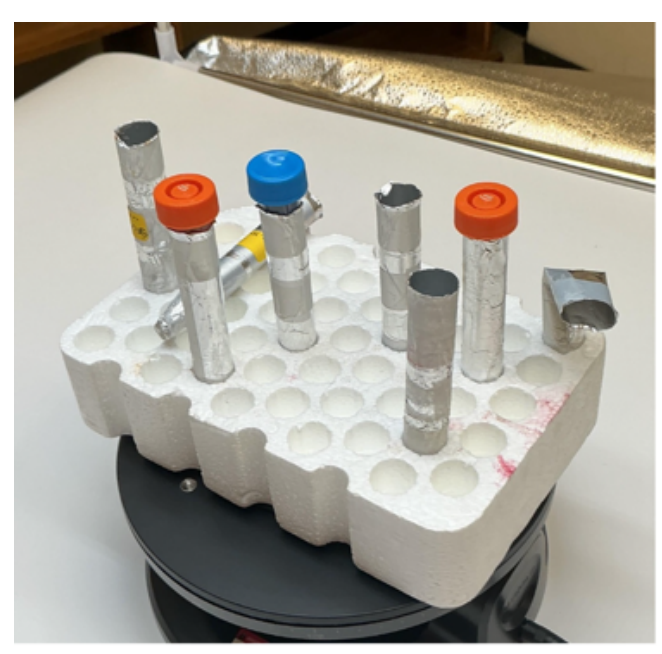

| Method | Shape | Texture | Color | Defects | All Traits |
|---|---|---|---|---|---|
| Photogram | 4.7 ± 0.8 | 4.5 ± 0.7 | 4.5 ± 1.0 | 4.6 ± 1.0 | 4.6 ± 0.9 |
| Gaussian | 4.7 ± 0.7 | 3.8 ± 1.3 | 4.4 ± 0.8 | 4.2 ± 1.0 | 4.3 ± 1.0 |
| LiDAR | 4.7 ± 0.8 | 3.9 ± 1.2 | 4.7 ± 0.6 | 4.5 ± 1.1 | 4.4 ± 1.0 |
| NeRF | 4.7 ± 0.8 | 3.8 ± 1.1 | 4.1 ± 0.8 | 4.5 ± 1.1 | 4.3 ± 1.0 |
| All Methods | 4.7 ± 0.7 | 4.0 ± 1.1 | 4.4 ± 0.8 | 4.4 ± 1.0 | |

**Table S-11**. **Object 11**. Unsprayed Mortar and Pestle

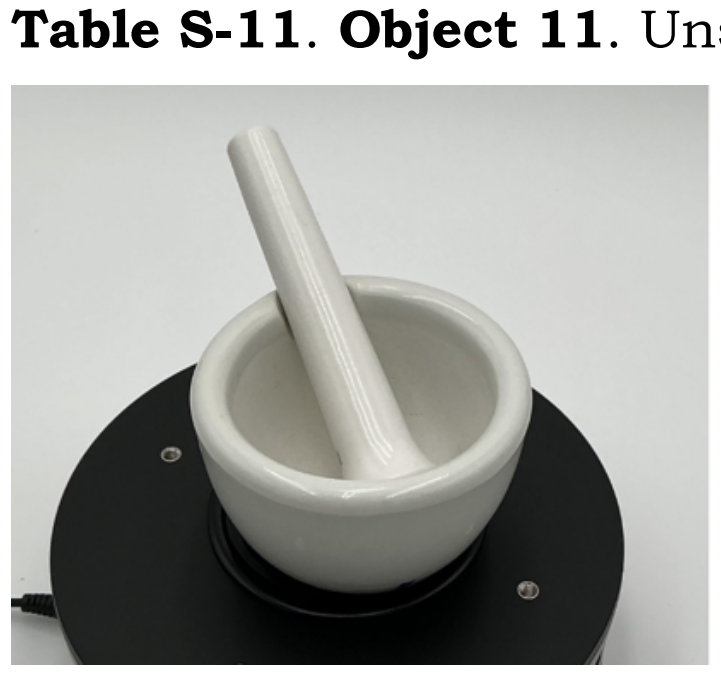

| Method | Shape | Texture | Color | Defects | All Traits |
|---|---|---|---|---|---|
| Photogram | 3.4 ± 1.2 | 1.8 ± 1.0 | 3.5 ± 1.2 | 2.4 ± 0.9 | 2.8 ± 1.3 |
| Gaussian | 2.5 ± 1.6 | 2.4 ± 1.1 | 3.4 ± 1.3 | 1.3 ± 0.9 | 2.4 ± 1.4 |
| LiDAR | 4.1 ± 1.2 | 3.2 ± 1.2 | 3.7 ± 1.0 | 3.5 ± 1.0 | 3.6 ± 1.1 |
| NeRF | 2.8 ± 1.3 | 2.5 ± 1.6 | 3.6 ± 1.5 | 2.2 ± 1.0 | 2.8 ± 1.4 |
| All Methods | 3.2 ± 1.4 | 2.5 ± 1.3 | 3.6 ± 1.2 | 2.4 ± 1.2 | |

**Table S-12**. **Object 12**. Sprayed Mortar and Pestle

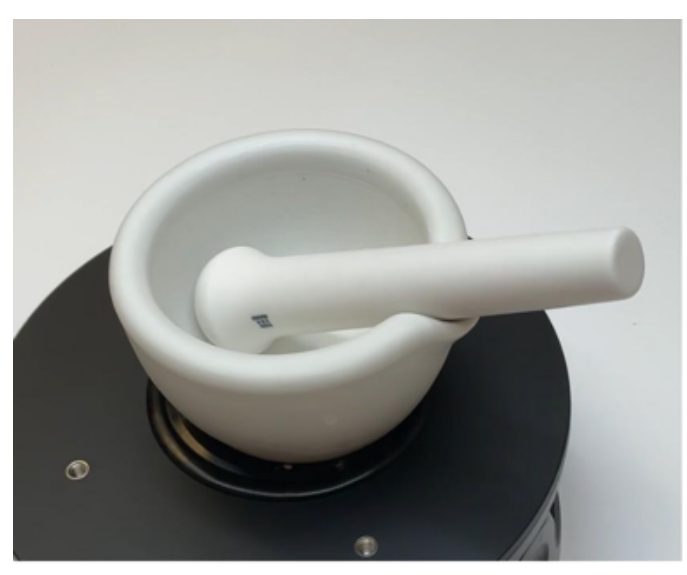

| Method | Shape | Texture | Color | Defects | All Traits |
|---|---|---|---|---|---|
| Photogram | 3.7 ± 1.2 | 2.6 ± 1.1 | 3.7 ± 1.0 | 3.3 ± 0.6 | 3.3 ± 1.1 |
| Gaussian | 2.0 ± 1.2 | 1.9 ± 1.0 | 3.4 ± 1.1 | 1.4 ± 0.3 | 2.1 ± 1.2 |
| LiDAR | 4.2 ± 1.1 | 3.5 ± 0.9 | 3.8 ± 1.0 | 3.7 ± 0.8 | 3.8 ± 0.9 |
| NeRF | 3.2 ± 1.2 | 2.9 ± 0.9 | 3.8 ± 0.9 | 2.7 ± 0.8 | 3.1 ± 1.0 |
| All Methods | 3.2 ± 1.4 | 2.7 ± 1.1 | 3.6 ± 1.0 | 2.8 ± 1.2 | |

**Supporting Table S-13**. Friedman analysis of statistically meaningful differences for each object (**1** - **12**) and visual attribute (shape, texture, color, defects) amongst the four different methods (photogrammetry, Gaussian Splatting, LiDAR, NERF). $p < 0.05$ for instances where the methods show statistically significant variances, are boldfaced, and were assessed using *post hoc* paired comparisons (see **Table S-14**).

| Object_ID | Attribute | p |
|---|---|---|
| **1** | **shape** | 3.7 E-06 |
| **1** | **texture** | 0.00015 |
| 1 | color | 0.063 |
| **1** | **defects** | 3.0 E-08 |
| **2** | **shape** | 2.6 E-07 |
| **2** | **texture** | 7.2 E-07 |
| **2** | **color** | 0.0097 |
| **2** | **defects** | 7.5 E-09 |
| 3 | shape | 0.60 |
| **3** | **texture** | 0.033 |
| 3 | color | 0.36 |
| 3 | defects | 0.11 |
| **4** | **shape** | 2.6 E-07 |
| **4** | **texture** | 4.2 E-05 |
| **4** | **color** | 0.015 |
| **4** | **defects** | 3.0 E-08 |
| **5** | **shape** | 0.0047 |
| **5** | **texture** | 0.014 |
| 5 | color | 0.098 |
| **5** | **defects** | 3.8 E-05 |
| 6 | shape | 0.84 |
| 6 | texture | 0.45 |
| 6 | color | 0.21 |
| 6 | defects | 0.66 |
| 7 | shape | 0.45 |
| **7** | **texture** | 0.0026 |
| 7 | color | 0.18 |
| 7 | defects | 0.62 |
| **8** | **shape** | 0.00015 |
| **8** | **texture** | 3.7 E-06 |
| **8** | **color** | 0.0047 |
| **8** | **defects** | 1.5 E-06 |
| 9 | shape | 0.44 |
| 9 | texture | 0.15 |
| 9 | color | 0.36 |
| 9 | defects | 0.36 |
| 10 | shape | 0.57 |
| **10** | **texture** | 0.04 |
| **10** | **color** | 0.015 |
| 10 | defects | 0.15 |
| **11** | **shape** | 0.00043 |
| **11** | **texture** | 0.016 |
| 11 | color | 0.84 |
| **11** | **defects** | 1.2 E-05 |
| **12** | **shape** | 2.5 E-07 |
| **12** | **texture** | 0.00063 |
| 12 | color | 0.080 |
| **12** | **defects** | 2.5 E-07 |

**Supporting Table S-14**. *Post hoc* paired comparisons between methods were performed on those objects and attributes that displayed statistically significant differences using Friedman's test (**Table S-13**). All tests are two-sided, where multiplicities are accounted for using Holm's method. p-values < 0.05 are considered statistically significant and, in those instances, the scanning method that achieved the better score is boldfaced.

| Object_ID | Attribute | Method 1 | Method 2 | p |
|---|---|---|---|---|
| **1** | shape | Photogram | **Gaussian** | 0.0089 |
| **1** | shape | Photogram | LiDAR | 0.40 |
| **1** | shape | Photogram | **NeRF** | 0.0040 |
| **1** | shape | **Gaussian** | LiDAR | 0.0054 |
| **1** | shape | Gaussian | NeRF | 0.11 |
| **1** | shape | LiDAR | **NeRF** | 0.0042 |
| **1** | texture | Photogram | **Gaussian** | 0.0072 |
| **1** | texture | Photogram | **LiDAR** | 0.0045 |
| **1** | texture | Photogram | **NeRF** | 0.0045 |
| **1** | texture | Gaussian | LiDAR | 0.61 |
| **1** | texture | Gaussian | NeRF | 0.45 |
| **1** | texture | LiDAR | NeRF | 0.21 |
| **1** | defects | Photogram | **Gaussian** | 0.0013 |
| **1** | defects | Photogram | **LiDAR** | 0.0013 |
| **1** | defects | Photogram | **NeRF** | 0.0011 |
| **1** | defects | **Gaussian** | LiDAR | 0.0075 |
| **1** | defects | Gaussian | NeRF | 0.059 |
| **1** | defects | LiDAR | **NeRF** | 0.0013 |
| **2** | shape | Photogram | **Gaussian** | 0.0049 |
| **2** | shape | Photogram | **LiDAR** | 0.038 |
| **2** | shape | Photogram | **NeRF** | 0.0023 |
| **2** | shape | **Gaussian** | LiDAR | 0.026 |
| **2** | shape | Gaussian | **NeRF** | 0.0019 |
| **2** | shape | LiDAR | **NeRF** | 0.0016 |
| **2** | texture | Photogram | **Gaussian** | 0.0087 |
| **2** | texture | Photogram | **LiDAR** | 0.0042 |
| **2** | texture | Photogram | **NeRF** | 0.0039 |
| **2** | texture | Gaussian | LiDAR | 0.82 |
| **2** | texture | Gaussian | **NeRF** | 0.0039 |
| **2** | texture | LiDAR | **NeRF** | 0.0087 |
| **2** | color | Photogram | Gaussian | 0.12 |
| **2** | color | Photogram | LiDAR | 0.32 |
| **2** | color | Photogram | NeRF | 0.054 |
| **2** | color | Gaussian | LiDAR | 0.77 |
| **2** | color | Gaussian | NeRF | 0.48 |
| **2** | color | LiDAR | NeRF | 0.48 |
| **2** | defects | Photogram | **Gaussian** | 0.0011 |
| **2** | defects | Photogram | **LiDAR** | 0.0011 |
| **2** | defects | Photogram | **NeRF** | 0.00070 |
| **2** | defects | Gaussian | LiDAR | 0.24 |
| **2** | defects | Gaussian | **NeRF** | 0.0011 |
| **2** | defects | LiDAR | **NeRF** | 0.0011 |
| **3** | texture | Photogram | Gaussian | 0.26 |
| **3** | texture | Photogram | LiDAR | 0.45 |
| **3** | texture | Photogram | NeRF | 0.82 |
| **3** | texture | Gaussian | LiDAR | 0.17 |
| **3** | texture | Gaussian | NeRF | 0.16 |
| **3** | texture | LiDAR | NeRF | 0.53 |
| **4** | shape | Photogram | **Gaussian** | 0.046 |
| **4** | shape | Photogram | LiDAR | 0.49 |
| **4** | shape | Photogram | **NeRF** | 0.0020 |
| **4** | shape | Gaussian | **LiDAR** | 0.011 |
| **4** | shape | Gaussian | **NeRF** | 0.0016 |
| **4** | shape | LiDAR | **NeRF** | 0.0020 |
| **4** | texture | Photogram | Gaussian | 0.18 |
| **4** | texture | Photogram | LiDAR | 0.23 |
| **4** | texture | Photogram | **NeRF** | 0.029 |
| **4** | texture | Gaussian | LiDAR | 0.029 |
| **4** | texture | Gaussian | **NeRF** | 0.0073 |
| **4** | texture | LiDAR | **NeRF** | 0.029 |
| **4** | color | Photogram | Gaussian | 0.61 |
| **4** | color | Photogram | LiDAR | 0.69 |
| **4** | color | Photogram | NeRF | 0.18 |
| **4** | color | Gaussian | LiDAR | 0.69 |
| **4** | color | Gaussian | NeRF | 0.11 |
| **4** | color | LiDAR | NeRF | 0.18 |
| **4** | defects | **Photogram** | Gaussian | 0.041 |
| **4** | defects | Photogram | **LiDAR** | 0.018 |
| **4** | defects | Photogram | **NeRF** | 0.0015 |
| **4** | defects | Gaussian | **LiDAR** | 0.0056 |
| **4** | defects | Gaussian | **NeRF** | 0.0015 |

| **4** | defects | LiDAR | **NeRF** | 0.0015 |
|---|---|---|---|---|
| **5** | shape | Photogram | Gaussian | 0.77 |
| **5** | shape | Photogram | LiDAR | 0.71 |
| **5** | shape | Photogram | NeRF | 0.17 |
| **5** | shape | Gaussian | LiDAR | 0.71 |
| **5** | shape | Gaussian | NeRF | 0.21 |
| **5** | shape | LiDAR | **NeRF** | 0.027 |
| **5** | texture | Photogram | Gaussian | 0.42 |
| **5** | texture | Photogram | LiDAR | 0.18 |
| **5** | texture | Photogram | NeRF | 0.20 |
| **5** | texture | Gaussian | LiDAR | 0.42 |
| **5** | texture | Gaussian | NeRF | 0.060 |
| **5** | texture | LiDAR | NeRF | 0.060 |
| **5** | defects | Photogram | Gaussian | 0.61 |
| **5** | defects | Photogram | LiDAR | 1 |
| **5** | defects | Photogram | **NeRF** | 0.0093 |
| **5** | defects | Gaussian | LiDAR | 1 |
| **5** | defects | Gaussian | **NeRF** | 0.0092 |
| **5** | defects | LiDAR | **NeRF** | 0.0084 |
| **7** | texture | **Photogram** | Gaussian | 0.019 |
| **7** | texture | Photogram | LiDAR | 0.64 |
| **7** | texture | Photogram | NeRF | 0.52 |
| **7** | texture | Gaussian | **LiDAR** | 0.019 |
| **7** | texture | Gaussian | **NeRF** | 0.037 |
| **7** | texture | LiDAR | NeRF | 0.58 |
| **8** | shape | Photogram | **Gaussian** | 0.017 |
| **8** | shape | Photogram | **LiDAR** | 0.039 |
| **8** | shape | Photogram | **NeRF** | 0.0058 |
| **8** | shape | Gaussian | LiDAR | 0.28 |
| **8** | shape | Gaussian | **NeRF** | 0.020 |
| **8** | shape | LiDAR | NeRF | 0.20 |
| **8** | texture | Photogram | **Gaussian** | 0.0057 |
| **8** | texture | Photogram | **LiDAR** | 0.0057 |
| **8** | texture | Photogram | **NeRF** | 0.0057 |
| **8** | texture | Gaussian | LiDAR | 0.11 |
| **8** | texture | Gaussian | NeRF | 0.43 |
| **8** | texture | LiDAR | NeRF | 0.11 |
| **8** | color | Photogram | **Gaussian** | 0.044 |
| **8** | color | Photogram | **LiDAR** | 0.044 |
| **8** | color | Photogram | NeRF | 0.23 |
| **8** | color | Gaussian | LiDAR | 0.50 |
| **8** | color | Gaussian | NeRF | 0.82 |
| **8** | color | LiDAR | NeRF | 0.50 |
| **8** | defects | Photogram | **Gaussian** | 0.0041 |
| **8** | defects | Photogram | **LiDAR** | 0.0026 |
| **8** | defects | Photogram | **NeRF** | 0.0041 |
| **8** | defects | Gaussian | **LiDAR** | 0.0027 |
| **8** | defects | Gaussian | **NeRF** | 0.0041 |
| **8** | defects | LiDAR | NeRF | 0.18 |
| **10** | texture | Photogram | Gaussian | 0.076 |
| **10** | texture | Photogram | LiDAR | 0.42 |
| **10** | texture | Photogram | NeRF | 0.18 |
| **10** | texture | Gaussian | LiDAR | 1 |
| **10** | texture | Gaussian | NeRF | 1 |
| **10** | texture | LiDAR | NeRF | 1 |
| **10** | color | Photogram | Gaussian | 0.97 |
| **10** | color | Photogram | LiDAR | 0.97 |
| **10** | color | Photogram | NeRF | 0.23 |
| **10** | color | Gaussian | LiDAR | 0.72 |
| **10** | color | Gaussian | NeRF | 0.21 |
| **10** | color | LiDAR | NeRF | 0.090 |
| **11** | shape | **Photogram** | Gaussian | 0.044 |
| **11** | shape | Photogram | LiDAR | 0.060 |
| **11** | shape | Photogram | NeRF | 0.11 |
| **11** | shape | Gaussian | **LiDAR** | 0.030 |
| **11** | shape | Gaussian | NeRF | 0.50 |
| **11** | shape | **LiDAR** | NeRF | 0.0082 |
| **11** | texture | Photogram | Gaussian | 0.53 |
| **11** | texture | Photogram | **LiDAR** | 0.0053 |
| **11** | texture | Photogram | NeRF | 0.53 |
| **11** | texture | Gaussian | LiDAR | 0.26 |
| **11** | texture | Gaussian | NeRF | 0.91 |
| **11** | texture | LiDAR | NeRF | 0.53 |
| **11** | defects | **Photogram** | Gaussian | 0.025 |
| **11** | defects | Photogram | **LiDAR** | 0.0065 |
| **11** | defects | Photogram | NeRF | 0.76 |
| **11** | defects | Gaussian | **LiDAR** | 0.00487 |
| **11** | defects | Gaussian | **NeRF** | 0.022 |

| **11** | defects | **LiDAR** | NeRF | 0.038 |
|---|---|---|---|---|
| **12** | shape | **Photogram** | Gaussian | 0.0024 |
| **12** | shape | Photogram | **LiDAR** | 0.017 |
| **12** | shape | Photogram | NeRF | 0.052 |
| **12** | shape | Gaussian | LiDAR | 0.0024 |
| **12** | shape | Gaussian | **NeRF** | 0.0074 |
| **12** | shape | **LiDAR** | NeRF | 0.0074 |
| **12** | texture | Photogram | Gaussian | 0.20 |
| **12** | texture | Photogram | LiDAR | 0.058 |
| **12** | texture | Photogram | NeRF | 0.41 |
| **12** | texture | Gaussian | **LiDAR** | 0.0055 |
| **12** | texture | Gaussian | NeRF | 0.090 |
| **12** | texture | LiDAR | NeRF | 0.20 |
| **12** | defects | **Photogram** | Gaussian | 0.0015 |
| **12** | defects | Photogram | LiDAR | 0.11 |
| **12** | defects | Photogram | NeRF | 0.11 |
| **12** | defects | Gaussian | **LiDAR** | 0.0019 |
| **12** | defects | Gaussian | **NeRF** | 0.0021 |